\documentclass[12pt]{iopart}
\usepackage{graphics}
\usepackage{graphicx}
\input{epsf}
\def\l{$\Lambda$}

\def\etal{et al.}

\def\eg{e.g.}

\def \lleq {\lower0.9ex\hbox{ $\buildrel < \over \sim$} ~}
\def \ggeq {\lower0.9ex\hbox{ $\buildrel > \over \sim$} ~}

\def\omt{\Omega_{\rm m}}
\def\om{\Omega_{0 {\rm m}}}
\def\l{\Lambda}

\def\spose#1{\hbox to 0pt{#1\hss}}
\def\simle{\mathrel{\spose{\lower 3pt\hbox{$\mathchar"218$}}
     \raise 2.0pt\hbox{$\mathchar"13C$}}}
\def\simge{\mathrel{\spose{\lower 3pt\hbox{$\mathchar"218$}}
     \raise 2.0pt\hbox{$\mathchar"13E$}}}
\def\apj{Astroph.~J.~}

\def\aj{Astron.~J.~}
\def\prl{Phys.~Rev.~Lett.~}
\def\prd{Phys.~Rev.~D~}

\def\beq{\begin{equation}}
\def\eeq{\end{equation}}
\def\ber{\begin{eqnarray}}
\def\eer{\end{eqnarray}}
\def\etal{{\it et al.}}

\def\n {\noindent}

\newcommand{\sq}{\lower.25ex\hbox{\large$\Box$}}

\begin{document}

\title{Can dark energy be decaying?}  
\author{
Ujjaini Alam ${}^{a}$, Varun Sahni${}^{a}$
and A. A. Starobinsky${}^{b}$}   

\address{${}^{a}$ Inter-University Centre for Astronomy \& Astrophysics,
Pun\'e 411 007, India}
\address{${}^{b}$ Landau Institute for Theoretical Physics,
119334 Moscow, Russia}
\date{\today}

\begin{abstract}
We explore the fate of the universe given the possibility that the
density associated with `dark energy' may decay slowly with time.
Decaying dark energy is modeled by a homogeneous scalar field which
couples minimally to gravity and whose potential has {\em at least
  one} local quadratic maximum. Dark energy decays as the scalar field
rolls down its potential, consequently the current acceleration epoch
is a transient.  We examine two models of decaying dark energy.  In
the first, the dark energy potential is modeled by an analytical form
which is generic close to the potential maximum. The second potential
is the cosine, which can become negative as the field evolves,
ensuring that a spatially flat universe collapses in the future.  We
examine the feasibility of both models using observations of high
redshift type Ia supernovae. A maximum likelihood analysis is used to
find allowed regions in the $\lbrace m, \phi_0\rbrace$ plane ($m$ is
the tachyon mass modulus and $\phi_0$ the initial scalar field value;
$m\sim H_0$ and $\phi_0\sim M_P$ by order of magnitude).  For the
first model, the time for the potential to drop to half its maximum
value is larger than $\sim 8$ Gyrs.  In the case of the cosine
potential, the time left until the universe collapses is always
greater than $\sim 18$ Gyrs (both estimates are presented for $\om =
0.3$, $m/H_0 \sim 1$, $H_0 \simeq 70$ km/sec/Mpc, and at the 95.4\%
confidence level).
\end{abstract}

\bigskip
\section{Introduction}

Observations of the luminosity distance to distant type Ia supernovae
\cite{SN1,SN2} supported by the discovery of CMB angular temperature
fluctuations on degree scales \cite{CMB} and measurements of the power
spectrum of galaxy clustering \cite{perc} convincingly show that our
Universe is (approximately) spatially flat, with $\sim 30\%$ of its
critical energy density in non-relativistic matter (cold dark matter
(CDM) and baryons) and the remaining $\sim 70\%$ in a smooth component
having a large negative pressure (`dark energy'). Dark energy is
clearly the most abundant form of matter in the Universe (in terms of
the effective energy density), yet both its nature and its
cosmological origin remain enigmatic at present. It is clear that the
observational properties of dark energy (DE) are tantalizingly close
to those of a cosmological constant `$\l$', yet dark energy {\em need
  not} be $\l$ exactly (see reviews \cite{ss99,bp02}).

An intriguing possibility, carrying far reaching consequences both for
the current Universe and its ultimate fate, is that the dark energy
may be {\em decaying}.  The possibility that dark energy could be
unstable is in fact suggested by the remarkable {\em qualitative}
analogy between the presence of dark energy today and the properties
of a different type of `dark energy' -- the inflaton field --
postulated in the inflationary scenario of the early Universe. This
analogy works in two ways. On one hand, the fact that a form of
matter having a large negative pressure dominates the universe today
makes it not unnatural that a similar form of matter (having $w < 0$)
could have dominated the universe in the distant past.  On the other
hand, since dark energy in the early Universe (the inflaton) was
unstable and decayed aeons ago, one might be tempted to ask whether
the nature of dark energy observed today will be any different.  In
this paper we address in detail the possibility that dark energy, like
the inflaton which preceded it, may be decaying.

Decaying dark energy (DDE) leads to numerous interesting possibilities
including the fact that the current epoch of cosmic acceleration could
be a transient which ends after the dark energy density has dropped to
sufficiently small values. Such a universe will clearly be very
different from the standard $\Lambda$CDM cosmology. For instance, a
DDE universe may not possess horizons which are characteristic of
$\Lambda$CDM as well as tracker-driven quintessence models (see
\cite{sahni02} for a brief review). (Interesting implications for the
future of our universe also arise in certain braneworld models, in
which the current accelerating regime is a transient between two
matter dominated epochs \cite{ss02a,as02,ss02b}.)

Within the context of quintessence models another interesting (though
more speculative) possibility is provided by potentials which are not
constrained to be positive but which can become negative for certain
values of $\phi$ \cite{ffkl02}.  These potentials are either bounded
from below, in which case $V(\phi)$ has one or more minima at which
$V(\phi_i) < 0$.  (A good example is furnished by the cosine potential
which will be examined in detail later in this paper.)  Alternatively
the potential becomes unbounded, in which case $V(\phi) \to -\infty$,
for some values of $\phi$.  Under the influence of a negative
potential, DDE steadily decreases and the universe expands at an
increasingly slower rate. Finally a stage is reached when the negative
DDE density {\em exactly cancels} the positive density of dark matter
(plus the curvature term, if present). At this point of time the
expansion of the universe stops ($H=0$) and thereafter the universe
begins to collapse ($H<0$). This paper examines two general classes of
potentials describing decaying dark energy in the light of recent high
redshift supernova observations. In the first class $V(\phi) = V_0 -
\frac{1}{2}m^2\phi^2$, and we evolve the field equations into the
future to determine the time taken for the potential to drop to half
its maximum value.  The second class of models involves the cosine
potential. In both cases we use the agreement between our model and
high redshift supernova data {\em in the past}, to extrapolate and
predict the behaviour of our model universe {\em in the future}.  As a
result, we find the range of model parameters allowed by the
supernovae data and determine: (i) the {\em minimum} time for the
universe to {\em stop accelerating}, and, (ii) the {\em minimum} time
for the universe to {\em collapse} in DDE models with negative
potentials.

\section{Decaying Dark Energy}

The decaying dark energy (DDE) model which we consider is analogous to
successful inflationary models in that it undergoes quasi-homogeneous
decay.  We model DDE by a scalar field $\phi$ (`quintessence') which
couples minimally to gravity and has one or more maxima in its
potential.  Near any one such local maximum (which for convenience we
assume is at $\phi=0$) the field can be {\em generically} modeled by
the potential
\beq \label{eq:V}
V(\phi)=V_0-{m^2\phi^2\over 2}~, ~~{\rm where} ~~ m^2 \equiv \vert
\frac{d^2V}{d\phi^2}\vert\bigg\vert_{\phi=0} ~.  
\eeq 
The modulus of the tachyonic rest-mass $m$ will be assumed $\sim H_0$
where $H_0$ is the Hubble constant (the slow-roll condition, $m\ll
H_0$, would result in the absence of any noticeable observational
effects for ultra-light scalars, making such fields virtually
indistinguishable from $\Lambda$CDM, at least within the accuracy of
present data).  It is known from observations that physical properties
of dark energy are close to those of a cosmological constant at
present.  This means that the variation in $\phi$ must be on time
scales of order $H_0^{-1}$ and not much faster. (In the opposite case
the equation of state $w$ would vary much too rapidly.) The equation
of motion then implies that $m\sim H_0$ which suggests that all
characteristic time-scales in our problem will be of order $H_0^{-1}$
or greater.  Therefore, in this model the initial value $\phi_0$ of
the scalar field $\phi$ and its change -- during the entire process of
dark energy decay -- are both expected to be of the order of $G^{-1}$.
(This follows quite simply from the Einstein equation $m^2\phi^2 \sim
\rho_{\rm crit} \sim H_0^2/G$.)  However, since cosmology is now a
precision science, we can hope to obtain much better {\em
  quantitative} results, beyond these simple qualitative estimates,
using current observational data. This is precisely what is done in
this paper.

We should at this stage mention that, in principle, decay mechanisms
other than those considered in this paper are possible. As an example
one should mention bubble-like decay via quantum tunneling in the case
when $V'' > 0$, or strongly inhomogeneous classical decay in the case
of a tachyonic mass large compared to $H_0$, however these
possibilities must be confronted with observational data other than
the high-z supernova data which we wish to consider, and will
therefore not be discussed any further by us in this paper.  It may
also be appropriate to mention that our model of decaying dark energy
does not introduce any direct non-gravitational coupling of dark
energy to non-relativistic matter. Significant coupling of dark energy
to dark matter may lead to large energy transfer from dark energy to
dark matter in the process of DDE decay. However this model of DDE,
though potentially interesting, lies somewhat beyond the scope of the
present paper and we do not consider it any further here. In our case,
what we call the DE decay is actually the transition of DE from the
potential-dominated regime to a regime in which the potential and
kinetic energies of the dark energy field $\phi$ become comparable.
\footnote{We also do not consider fields which couple non-minimally to
  gravity.  Such a coupling, if small, would result in an
  insignificant change in the effective mass parameter $m$ for time
  intervals of the order of $H_0^{-1}$ which we consider in this
  paper. Large couplings would lead to a value of the effective
  Brans-Dicke parameter $\omega$ which violates the lower bounds on
  this parameter set by Solar system tests of Einstein gravity.}

The field equations governing the behaviour of the scalar field as it
rolls down its potential in a spatially flat universe are
\ber\label{eq:diffeqn}
{\ddot \phi} &+& 3H{\dot\phi} + V'(\phi) = 0~,\nonumber\\
H^2 &=& \frac{8\pi G}{3}\left (\rho_{m} + \rho_\phi\right )~,
\eer
where $V(\phi)$ is given by (\ref{eq:V}) and 
\beq
\rho_\phi = V(\phi) + \frac{1}{2}{\dot\phi}^2~.
\eeq
At early times ($H_0t\ll 1$), the universe was matter-dominated. 
Assuming $\rho_{m} \gg \rho_\phi$ in (\ref{eq:diffeqn}) we obtain the
following exact solution for $\phi(t)$
during the matter dominated regime:  
\beq\label{eq:exact}
\phi(t) = \phi_0\frac{\sinh{mt}}{mt}~.
\eeq
Eq (\ref{eq:exact}) determines the initial conditions for $\phi$ and
$\dot\phi$. At very early times ($mt \ll 1$) the very large damping
experienced by the scalar field ensures that it remains close to its
initial value at $\phi_0$.  One should note that this initial value
must be sufficiently small, $\phi_0 \lleq \sqrt{3/8\pi G}$, in order
that DE have sufficiently large negative pressure during the present
epoch.  This is clearly demonstrated in the next section when we
compare this model of dark energy with supernova observations.

Since the point $\phi=0,~\dot\phi =0$ is a saddle point for
homogeneous solutions of the scalar field equation with the potential
(\ref{eq:V}), there exists a set of non-zero measure of generic
solutions which remain sufficiently close to our solution up until the
present epoch. (It is sufficient for this purpose that a large initial
kinetic energy -- if at all it existed -- is redshifted by $z \sim 3$
so that the field settles on the trajectory described by
(\ref{eq:exact}) by that redshift.  Since the kinetic energy of the
scalar field decays as $\propto a^{-6}$ during the regime when $\dot
\phi^2 \gg m^2\phi^2$ this requirement is easily satisfied in
practice.)  Of course, one may with good reason ask as to whether our
set of initial conditions on $\lbrace\phi, {\dot\phi}\rbrace$ is not
too small. The answer to this question depends entirely on the
(unknown) behaviour of present dark energy at large temperatures and
curvatures where the form of the potential (\ref{eq:V}) need not be
valid. For this reason, we will not discuss it any further here.
However it should be noted that, generally speaking, the probability
of such initial conditions will be strongly enhanced if there are {\em
  many} maxima in the DE potential. This occurs, for instance, in the
DE model with $V(\phi)=V_0\cos (\phi/f)+V_1$
\cite{fhsw95,wf00,choi,ngwilt} which will be discussed by us later in
this paper (in the inflationary context, such a model is called
`natural inflation').

In the next section, we numerically integrate the system of equations
(\ref{eq:diffeqn}) with the potential (\ref{eq:V}), and determine the
scale factor $a(t)$ and the Hubble parameter $H(z)\equiv \dot a/a$ as
functions of the initial field displacement $\phi_0$ for specific
values of $m$ and $\om$.  Our model is confronted against high
redshift type Ia supernova observations through its luminosity
distance \beq\label{eq:lumdis} D_L = c\int\frac{dz}{H(z)}~, \eeq which
is used to place constraints on the free parameters of the model,
namely $m$ and $\phi_0$ ($V_0$ is not a free parameter since it is
uniquely determined by $m$, $\phi_0$, $\om$ and the current value of
the Hubble parameter -- $H_0$).

Having determined the permitted range of parameters values, we shall
proceed to determine the future evolution of our model universe.  As
emphasized in \cite{st00}, reliable future predictions are only
possible for finite intervals of time. Previous predictions (of the
kind `our Universe will keep expanding and not encounter a
singularity') had a depth of approximately $20$ Gyrs \cite{st00} (see
\cite{kl02,ft02,kl03} for similar estimates and \cite{as02} for the
braneworld context).  It is entirely reasonable to use the form
(\ref{eq:V}) up to the point when $m^2\phi^2=V_0$.  This occurs when
the potential has declined to half its maximum value. In the next
section we shall determine the minimal period of time for the
potential to reach its half-way mark (this time could be referred to
as the `half-decay time of dark energy'). Note that this estimate is
rather robust since it depends upon the fairly general form of the
potential near its maximum value given by (\ref{eq:V}) and not on any
other details of dark energy.

In Sec. 3, we also consider a potential which can become negative. The
form of this potential is assumed to be \cite{fhsw95,choi,ngwilt}:
\beq\label{eq:cos} V=V_0\cos{{\phi\over f}},~~~f= \frac{\sqrt V_0}{m}
\eeq where the value of $f$ is chosen in such a way that the potential
(\ref{eq:cos}) coincides with (\ref{eq:V}) at small $\phi$. Of course,
the hypothesis that $V$ may acquire negative values is speculative
since it does not follow from any of the current observational data.
However, such potentials often arise in supergravity and M-theory
models (see, e.g., \cite{choi,kl02,kl03}). For this potential, we
calculate: (i) the minimal extent of the current acceleration epoch,
and (ii) the time elapsed before the universe collapses. (We should
draw the readers attention to the fact that the collapse of the
universe is a generic property of flat cosmological models with
negative potentials \cite{ffkl02}.) We will not however consider the
subsequent period of contraction in any detail since its exact
duration strongly depends both upon the behaviour of $V$ in the region
$V<0$ and upon the properties of dark matter at high energies. We end
with section 4 which contains a summary of our results and a
discussion.

\section{Methodology and Results}

We constrain the parameter space of our cosmology by requiring that
our dark energy model provides a good fit to type Ia supernova data.
For this purpose we use the 54 SNe Ia from the primary `fit C' of the
Supernova Cosmology Project, which includes 16 low redshift
Calan-Tololo SNe \cite{SN1}. Fit C is a subsample of a total of 60 SNe
of which six are excluded as outliers: two low redshift SNe due to
suspected reddening and four high redshift SNe which are excluded due
to atypical light curves. The measured quantity in this data, the
bolometric magnitude $m_B$, is related to the luminosity distance and
therefore the cosmological parameters by the following equation

\beq\label{eq:mz}
m_B = {\cal M}+ 5 \ {\rm log}_{10} D_L(z;\om, m, \phi_0)\,\,,
\eeq

\n where $D_L=H_0 d_L$ is the Hubble-parameter-free luminosity
distance and $ {\cal M}= M_B + 25 -5 \ {\rm log}_{10} H_0$ is the
Hubble-parameter-free absolute magnitude.  We shall assume that the
SNe measurements come with uncorrelated Gaussian errors in which case
the likelihood function is given by the chi-squared distribution with
$N-n$ degrees of freedom: ${\cal L} \propto \exp{(-\chi^2/2)}$. (In
our case, $N = 54$ and $n=3$).

The $\chi^2$-statistic is defined as
\beq
\chi^2=\sum_{i=1}^{54} \left (\frac{m_i^{\rm eff}-
m(z_i)}{\sigma_{m_i}}\right)^2 \,\,,
\eeq
\n where $m_i^{\rm eff}$ is the effective B-band magnitude of the i-th
supernova obtained after correcting the observed magnitude at redshift
$z$ for the supernova width-luminosity relation, $\sigma_{m_i}$ is the
error in magnitude at redshift $z$, and $m(z_i)$ is the apparent
magnitude of the $i$-th supernova in our cosmological model.

For a given value of $\om$, the parameters to be estimated in our
cosmology are ${\cal M}$, $m$ and $\phi_0$. For all practical
purposes, the quantity ${\cal M}$ is a statistical nuisance parameter,
and we marginalize over it (assuming a uniform prior) when we
determine $m, ~\phi_0$. We perform the minimization while evolving the
scalar field according to the equations of motion (\ref{eq:diffeqn}).
For this purpose, we shall find it convenient to measure the parameter
$m$ in units of the quantity $\sqrt{8 \pi G V_0/3}$ and the parameter
$\phi_0$ in units of the reduced Planck mass $\tilde{M}_P=\sqrt{3/8
  \pi G}$.  (In these units we constrain the value of $m$ using the
prior $m \geq 0.1$. This condition is set to minimize numerical
uncertainties and to avoid models which are virtually
indistinguishable from $\Lambda$CDM).

\begin{figure*}
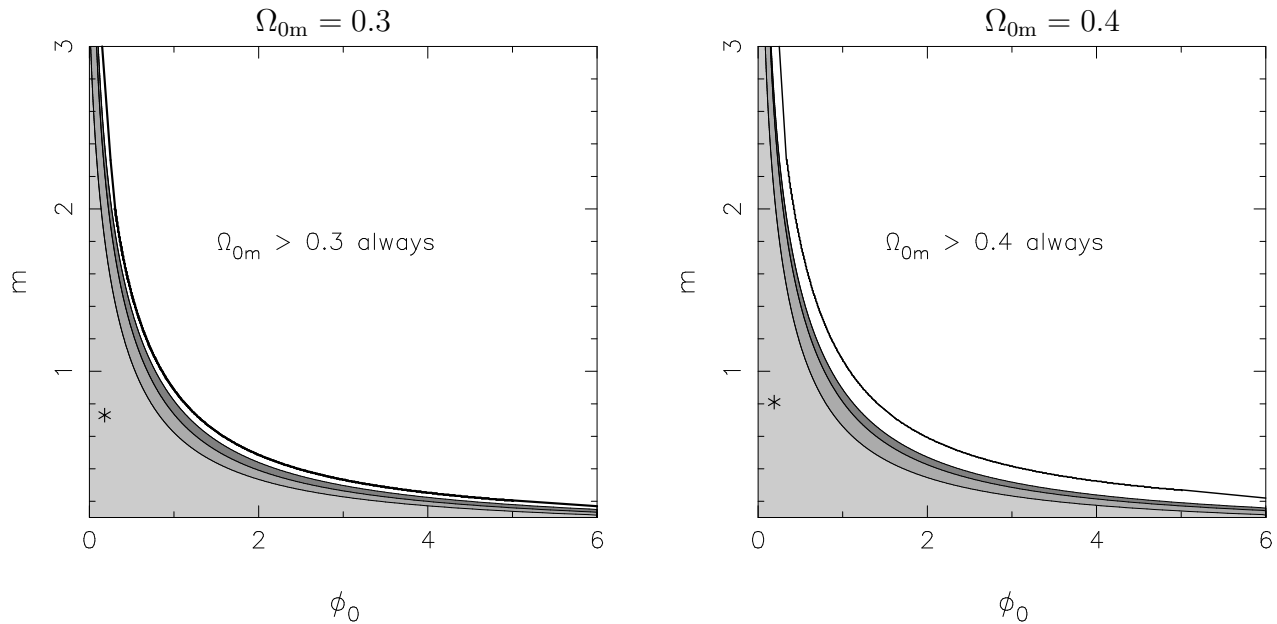
 
\centering
\begin{center}
\vspace{-0.05in}
\centerline{\mbox{\hspace{1.0in} $\om=0.3$ \hspace{3.0in} $\om=0.4$}}
\vspace{-0.42in}
$\begin{array}{c@{\hspace{0.4in}}c}
\multicolumn{1}{l}{\mbox{}} &
\multicolumn{1}{l}{\mbox{}} \\ [0.0cm]
\epsfxsize=3.1in
\epsffile{plotfull.epsi} &  
\epsfxsize=3.1in
\epsffile{plotfull04.epsi} \\  
\end{array}$
\end{center}
\caption{\small Confidence levels at $68.3\%$ (light grey inner contour) 
$95.4\%$ (medium grey contour) and $99.73\%$ (dark grey outer
contour) are shown in the $m$-$\phi_0$ plane for the potential
$V(\phi)=V_0-\frac{1}{2}m^2\phi^2$. In the left panel, the present
value of the matter density is $\om=0.3$, and in the right panel it is
$\om=0.4$. The mass $m$ is measured in units of $\sqrt{8 \pi G
  V_0/3}$, and the initial field value, $\phi_0$, is in units of the
reduced Planck mass $\tilde{M}_P=\sqrt{3/8 \pi G}$.  The best-fit
point in each plot is marked by a star.  The $\chi^2$ per degree of
freedom at the best-fit is $1.053$ for $\om=0.3$ and $1.049$ for
$\om=0.4$. For both figures, in the region to the right of the thick
solid line, parameter values are such that the matter density never
reaches its present value, hence this region is disallowed by
observations.  }
\label{fig:msq_full}
\end{figure*}

\begin{figure*}
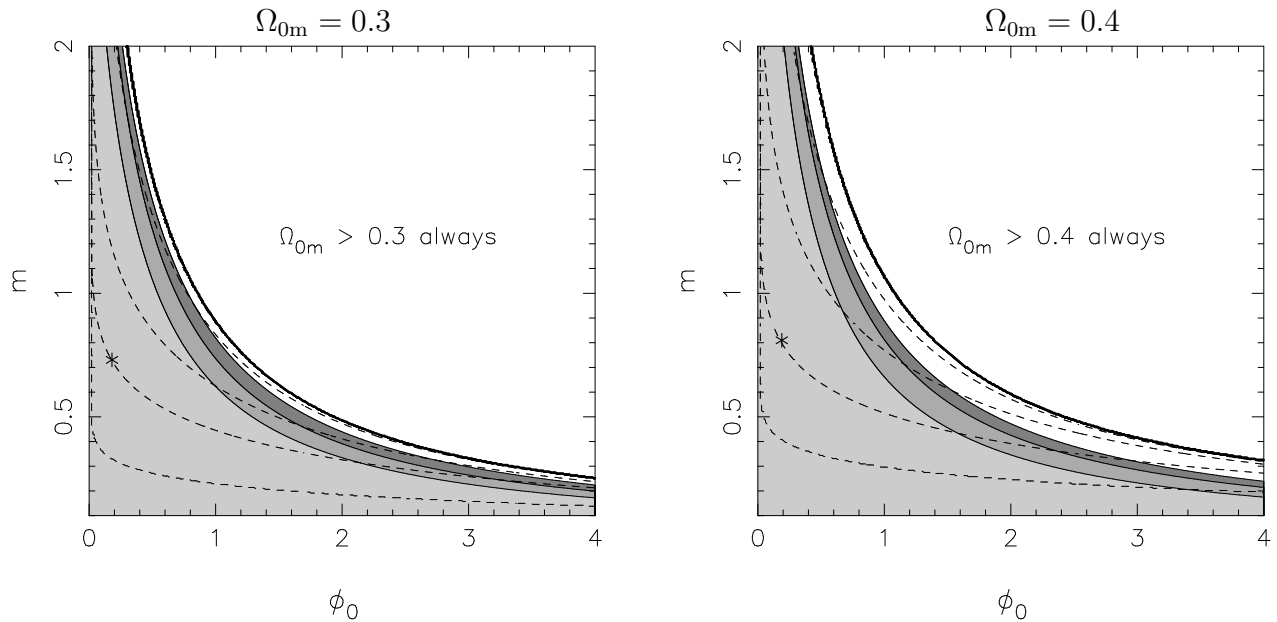
 
\centering
\begin{center}
\vspace{-0.05in}
\centerline{\mbox{{\hspace{1.0in} $\om=0.3$ \hspace{3.0in} $\om=0.4$}}}
\vspace{-0.42in}
$\begin{array}{c@{\hspace{0.4in}}c}
\multicolumn{1}{l}{\mbox{}} &
\multicolumn{1}{l}{\mbox{}} \\ [0.0cm]
\epsfxsize=3.1in
\epsffile{plot.epsi} &  
\epsfxsize=3.1in
\epsffile{plot04.epsi} \\  
\end{array}$
\end{center}
\caption{\small 
A magnified part of the Fig. \ref{fig:msq_full} with (dashed) lines of
constant $\Delta T_{1/2}$ added.  $\Delta T_{1/2}$ is the time,
measured from the present epoch, to when the DDE potential has dropped
to half its maximum value: $V(\phi)=V_0/2$. The values of $\Delta
T_{1/2}$ for the dashed curves (from top to bottom) are listed in
Table~\ref{tab:half} (from left to right). For both $\om=0.3$ and
$\om=0.4$, the minimum time elapsed before the potential drops to half
its maximum value is $\Delta T_{1/2} \simeq 0.6 H_0^{-1} \simeq 8 \ 
{\rm Gyrs}$($H_0=70 \ {\rm km/s/Mpc}$) at the $95.4 \%$ confidence
level.  In the region to the right of the thick solid curve, parameter
values are such that the matter density never reaches its present
value.  This region is therefore disallowed by observations.  }
\label{fig:msq}
\end{figure*}

\begin{figure*}
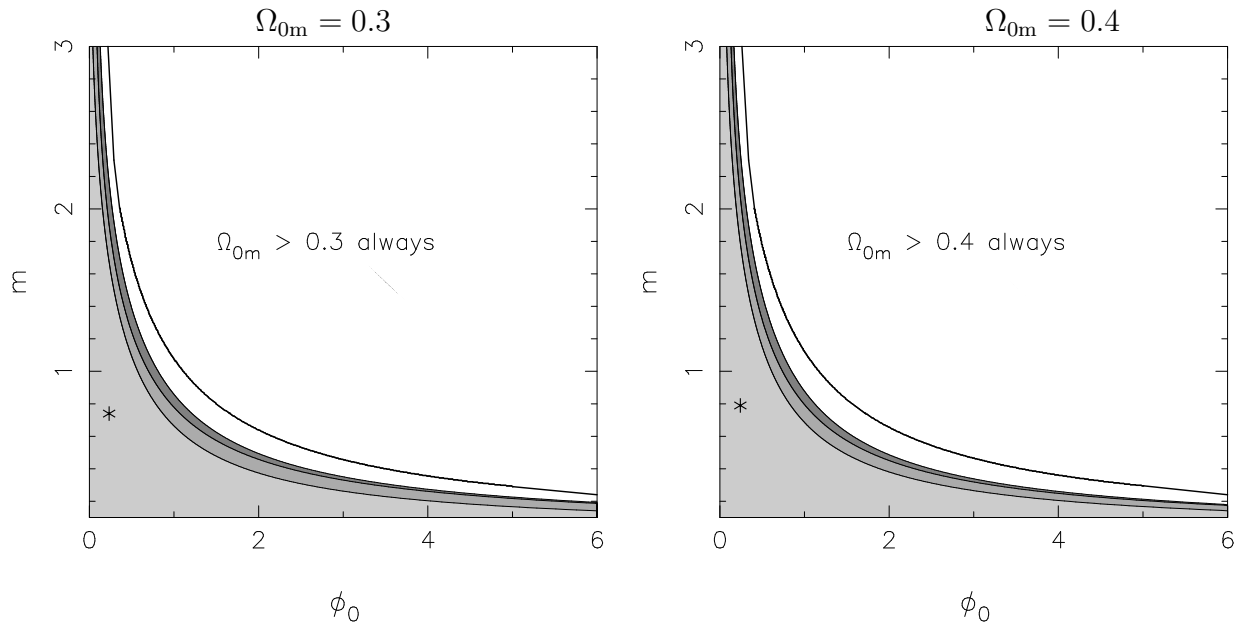
 
\centering
\begin{center}
\vspace{-0.05in}
\centerline{\mbox{{\hspace{1.0in} $\om=0.3$ \hspace{3.0in} $\om=0.4$}}}
\vspace{-0.42in}
$\begin{array}{c@{\hspace{0.2in}}c}
\multicolumn{1}{l}{\mbox{}} &
\multicolumn{1}{l}{\mbox{}} \\ [0.0cm]
\epsfxsize=3.1in
\epsffile{plotfull_cos.epsi} &  
\epsfxsize=3.1in
\epsffile{plotfull04_cos.epsi} \\  
\end{array}$
\end{center}
\caption{\small 
Confidence levels at $68.3\%$ (light grey inner contour) $95.4\%$
(medium grey contour) and $99.73\%$ (dark grey outer contour) are
shown in the $m$-$\phi_0$ plane for the potential $V(\phi)=V_0 {\rm
  cos}(m \phi/\sqrt{V_0})$.  Here $m$ is in units of $\sqrt{8 \pi G
  V_0/3}$, and $\phi_0$ is in units of the reduced Planck mass
$\tilde{M}_P=\sqrt{3/8 \pi G}$. The best-fit point in each plot is
marked as a star. The $\chi^2$ per degree of freedom at the best-fit
is $1.050$ for $\om=0.3$ and $1.047$ for $\om=0.4$. For both figures,
in the region to the right of the thick solid line, parameter values
are such that the matter density never reaches the present value,
hence this region is disallowed by observations.  }
\label{fig:cos_full}
\end{figure*}

\begin{figure*}
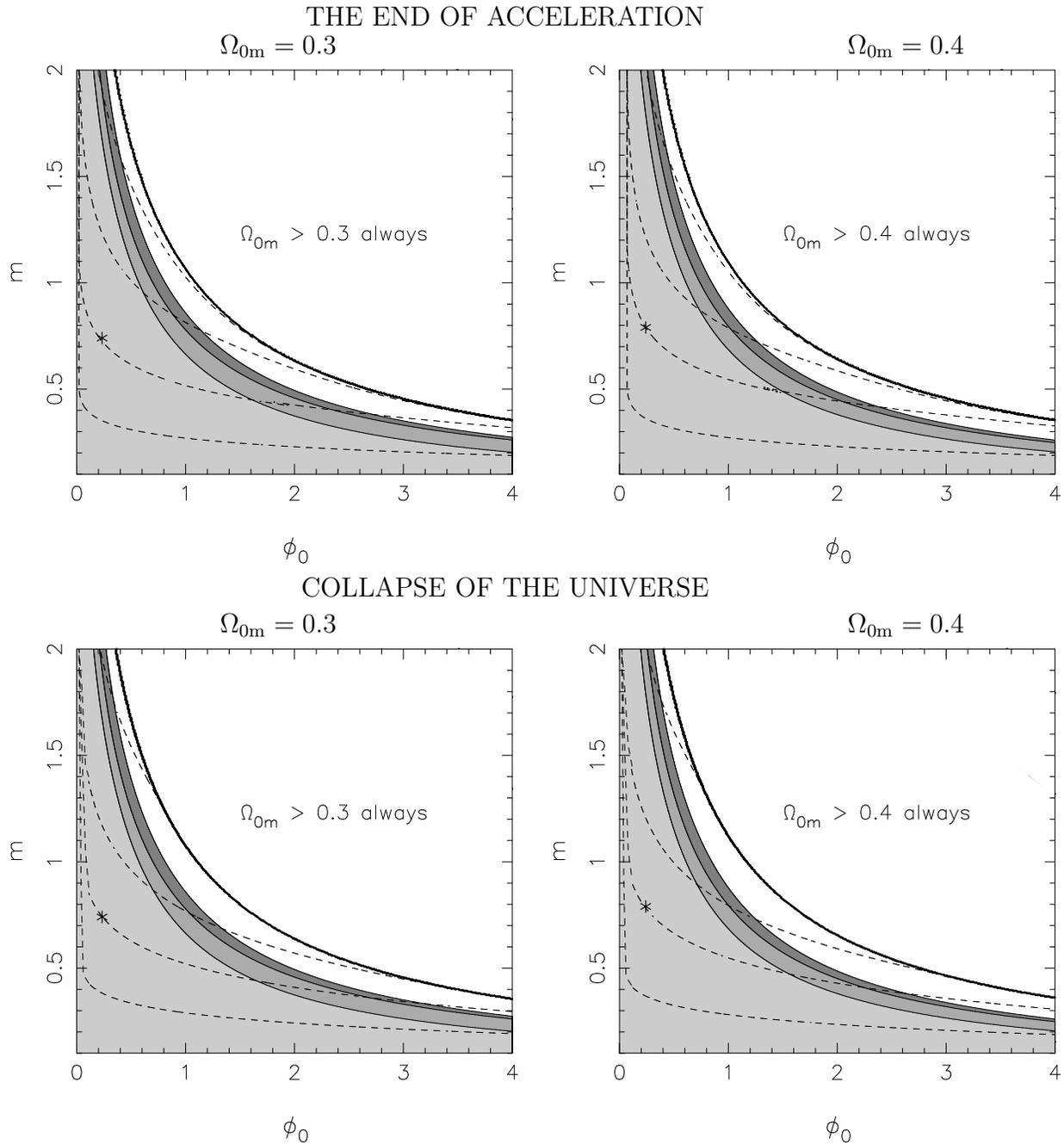
 
\centering
\begin{center}
\vspace{-0.05in}
\centerline{\mbox{THE END OF ACCELERATION}}
\vspace{-0.05in}
{\mbox{{\hspace{1.0in} $\om=0.3$ \hspace{3.0in} $\om=0.4$}}}
$\begin{array}{c@{\hspace{0.2in}}c}
\multicolumn{1}{l}{\mbox{}} &
\multicolumn{1}{l}{\mbox{}} \\ [-0.20in]
\epsfxsize=3.1in
\epsffile{plotq_cos.epsi} &  
\epsfxsize=3.1in
\epsffile{plotq04_cos.epsi} \\
\end{array}$
{\mbox{COLLAPSE OF THE UNIVERSE}}
\vspace{-0.05in}
{\mbox{{\hspace{1.0in} $\om=0.3$ \hspace{3.0in} $\om=0.4$}}}
$\begin{array}{c@{\hspace{0.2in}}c}
\multicolumn{1}{l}{\mbox{}} &
\multicolumn{1}{l}{\mbox{}} \\ [-0.15in]
\epsfxsize=3.1in
\epsffile{plot_cos.epsi} &  
\epsfxsize=3.1in
\epsffile{plot04_cos.epsi} \\  
\end{array}$
\end{center}
\vspace{-0.27in}
\caption{\small 
A magnified part of the Fig. \ref{fig:cos_full} with (dashed) lines of
constant $\Delta T_{\rm end}$ (upper panel) and constant $\Delta
T_{\rm coll}$ (lower panel) added.  In the upper panels, the time
$\Delta T_{\rm end}$ is measured from the present epoch to when the
universe stops accelerating: $q(t_0+\Delta T_{\rm end})=0$.  The
values of $\Delta T_{\rm end}$ for the dashed curves (from top to
bottom) are listed in Table~\ref{tab:collapse} (from left to right).
For both $\om=0.3$ and $\om=0.4$, the minimum time taken for the
deceleration parameter to rise to zero is $\Delta T_{\rm end} \simeq
0.7 H_0^{-1} \simeq 10 \ {\rm Gyrs}$ (at the $95.4 \%$ confidence
level). For the lower panel, the values of $\Delta T_{\rm coll}$ for
the dashed curves (from top to bottom) are listed in
Table~\ref{tab:collapse} (from left to right). For both $\om=0.3$ and
$\om=0.4$, the minimum time to collapse is $\Delta T_{\rm coll} \simeq
1.3 H_0^{-1} \simeq 18 \ {\rm Gyrs}$ at the $95.4 \%$ confidence level
(we assume $H_0=70 \ {\rm km/s/Mpc}$).  In the region to the right of
the thick solid curve the matter density never reaches its present
value of $\om = 0.3$ (left panel) and $\om = 0.4$ (right panel),
therefore this region is disallowed by observations.  }
\label{fig:cos_coll}
\end{figure*}

\begin{figure*}
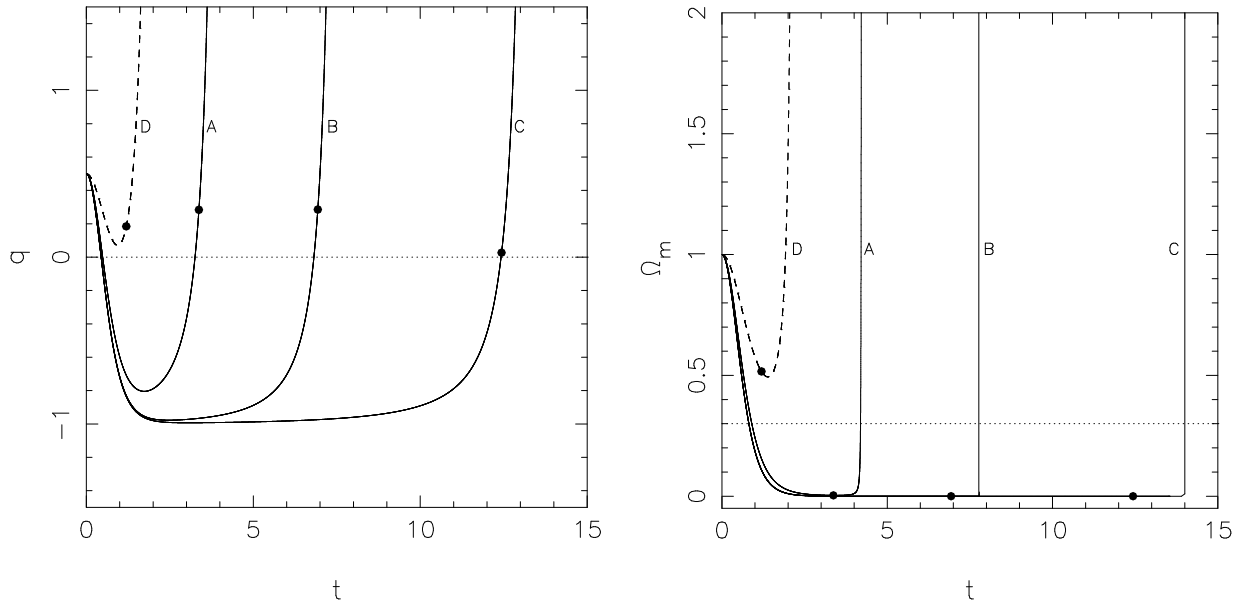
 
\centering
\begin{center}
$\begin{array}{c@{\hspace{0.2in}}c}
\multicolumn{1}{l}{\mbox{}} &
\multicolumn{1}{l}{\mbox{}} \\ [0.0cm]
\epsfxsize=3.1in
\epsffile{decel.epsi} & 
\epsfxsize=3.1in
\epsffile{dens.epsi} \\
\end{array}$
\end{center}
\caption{\small 
The evolution of the deceleration parameter $q$ and the matter density
$\omt$ is shown for four different DDE models corresponding to
different choices of $m$ and $\phi_0$ in the DDE potential
$V(\phi)=V_0 {\rm cos}(m \phi/\sqrt{V_0})$ ($\om = 0.3$).  Time $t$ is
in units of $\sqrt{3/8\pi G V_0}$.  The models have parameter values:
$m=1.0, \phi_0=0.6$ (A), $m=1.0, \phi_0=0.2$ (B), $m=0.74,
\phi_0=0.23$ (C), $m=1.0, \phi_0=1.2$ (D).  Models A,B,C are allowed
by supernova observations at the $95.4\%$ confidence level.  The
dashed line D in both panels shows the time evolution of $q$ and
$\omt$ for a DDE model with $m=1.0, \phi_0=1.2$.  This model is
disallowed by observations since the matter density always remains
larger than $0.3$ (see figure \ref{fig:cos_coll}).  The horizontal
dotted line in the left panel ($q=0$) divides this panel into two
regions. In the upper region $q > 0$ and the universe decelerates,
whereas $q < 0$ in the lower region in which the universe accelerates.
The points of intersection of $q=0$ with A,B,C show the commencement
and end of the acceleration epoch in these models.  The horizontal
dotted line in the right panel marks the present epoch when $\om=0.3$.
The solid circles in both left and right panels show the epoch when
the potential energy of the scalar field falls to zero. Note that this
occurs {\em after} the universe stops accelerating.  }
\label{fig:evolve}
\end{figure*}

In Figure~\ref{fig:msq_full}, we show the confidence levels in the
$m-\phi_0$ plane for the first potential described by Eq~(\ref{eq:V}).
We consider two cosmological models in which the current value of
$\om$ is $0.3$ and $0.4$ respectively. (The results for the two cases
differ marginally.) We see that an enormous region of parameter space
which corresponds to large values of $m$ \& $\phi_0$ is disallowed in
both models.  The reason for this is simple, for large values $m \gg
1$, $\phi_0 \gg 1$, the scalar field potential decreases much too
rapidly and the matter density never gets to reach its present value
of $\om = 0.3$ (left panel) or $\om = 0.4$ (left panel).  Also, from
the confidence levels we see that at higher values of $m$ ($\phi_0$),
the model provides a good fit to SNe data {\em only if} the
corresponding value of $\phi_0$ ($m$) is very small.  (It should be
noted that for asymptotically small values of either $m$ or $\phi_0$
the DDE model becomes virtually indistinguishable from $\l$CDM.)  From
the value of the $\chi^2$ at best-fit we see that higher values of
$\om$ seem to be slightly favoured by models of DDE. For $\om=0.3$,
$\chi^2_{\rm dof}=1.053$ for the best-fit model with $m=0.73,
\phi_0=0.18$, while $\l$CDM (for identical $\om$) has $\chi^2_{\rm
  dof}=1.054$. For $\om=0.4$, $\chi^2_{\rm dof}=1.049$ for the
best-fit model $m=0.81, \phi_0=0.19$, while $\chi^2_{\rm dof}(\l{\rm
  CDM})=1.058$ (for identical $\om$).  For lower values of $\om$,
$\eg$ $\om=0.2$, $\l$CDM is marginally favoured over the best-fit DDE.

In Figure~\ref{fig:msq}, we examine this potential in greater detail
by restricting ourselves to a smaller (and more probable) region in
parameter space.  In this region we show the lines of constant $\Delta
T_{1/2}$, where $\Delta T_{1/2}$ is the time taken, as measured from
the present epoch, for the potential to drop to half its maximum
value: $V\lbrack\phi(t_0+\Delta T_{1/2})\rbrack=V_0/2$.  We see that,
at the $95.4\%$ confidence level, the minimum time at which this
happens is $\Delta T_{1/2} \simeq 8$ Gyrs (if $H_0 \simeq 70$
km/sec/Mpc).

In Figure~\ref{fig:cos_full}, we show the confidence levels in the
$m-\phi_0$ plane for the second potential, described by
Eq~(\ref{eq:cos}).  The results in this case are qualitatively similar
to those for the first potential.  Here too, a large region of
parameter space corresponding to higher values of $m$ and $\phi_0$ is
disallowed because for these parameter values the matter density never
reaches its present day value. Also, at higher values of $m$
($\phi_0$), the model provides a good fit to SNe data only if the
complementary parameter $\phi_0$ ($m$) is very small. Again, for
asymptotically small values of either $m$ or $\phi_0$ this decaying
dark energy model becomes virtually indistinguishable from $\l$CDM.
For $\om=0.3$, we have $\chi^2_{\rm dof}=1.050$ at the best-fit point
of $m=0.74, \phi_0=0.23$, and for $\om=0.4$, $\chi^2_{\rm dof}=1.047$
for the best-fit model having $m=0.79, \phi_0=0.24$. As in the
previous analysis, $\l$CDM is marginally preferred over the best-fit
DDE for $\om < 0.3$.

\begin{table*}
\centering
\caption{Time taken for the potential (\ref{eq:V})
to drop to half its maximum value.}
\label{tab:half}
\begin{center}
\begin{tabular}{c|cccc} \hline
$\om$  & \centre{4}{$\Delta T_{1/2}$ (Gyr)} \\\hline
$0.3$ & $8.3$ & $42.0$ & $139.7$ & $712.8$ \\
$0.4$ & $8.3$ & $34.9$ & $153.6$ & $712.8$ \\\hline
\end{tabular}
\end{center}
\end{table*}

\begin{table*}
\centering
\caption{Time until the end of acceleration, $\Delta T_{\rm end}$.
Time until collapse, $\Delta T_{\rm coll}$.}
\label{tab:collapse}
\begin{center}
\begin{tabular}{c|cccc|cccc} \hline
$\om$  & \centre{4}{$\Delta T_{\rm end}$ (Gyr)} &  \centre{4}{$\Delta T_{\rm coll}$ (Gyr)} \\\hline
$0.3$ & $9.8$ & $34.9$ & $125.7$ & $698.4$ & $18.2$ & $55.9$ & $153.6$ & $698.4$ \\
$0.4$ & $9.8$ & $40.5$ & $111.7$ & $698.4$ & $18.2$ & $48.9$ & $125.7$ & $698.4$ \\\hline
\end{tabular}
\end{center}
\end{table*}

In Figure~\ref{fig:cos_coll}, we examine the results for this
potential more closely by focusing on a smaller region in parameter
space.  In this region we show the lines of constant $\Delta T_{\rm
  end}$, which is the time left until the universe ceases to
accelerate. (In terms of the deceleration parameter, $q(t_0 + \Delta
T_{\rm end}) = 0$.)  We see that, at the $95.4\%$ confidence level,
the universe will continue accelerating {\em for at least} $\sim 10$
Gyrs. We also plot the lines of constant $\Delta T_{\rm coll}$, which
is the time left until the universe collapses: $H(t_0+\Delta T_{\rm
  coll}) = 0$.  We find that the time to collapse is significantly
larger than the time to the end of acceleration. At the $95.4\%$
confidence level the lower bound to collapse is given by $\Delta
T_{\rm coll} \sim 18$ Gyrs for $m \lleq 2$.

In Figure~\ref{fig:evolve} we show the evolution of the deceleration
parameter $q$ and the matter density $\omt$ for DDE driven by the
cosine potential.  We see that models allowed by the SNe data show a
fairly large spread in the time span during which the universe
accelerates. Although the time when the universe starts to accelerate
is fairly close for different parameter values, the future epoch when
the universe stops accelerating, $q(t_0+\Delta T_{\rm end}) = 0$,
varies widely between models.  The same can be said for the time when
the universe collapses, which also varies significantly between
models.  Interestingly, for most allowed models, $\omt$ drops to
exceedingly small values before the potential becomes sufficiently
negative to initiate collapse.

\section{Discussion and Conclusions}

In this paper we have examined the possibility that the dark energy
responsible for the acceleration of the universe decays with time.
Models of decaying dark energy have recently been proposed in
connection with supergravity and M-theory \cite{kl02,kl03}.  Our
analysis of decaying dark energy is carried out for two potentials.
The first, defined in Eq. (\ref{eq:V}), is a faithful local
representation of a potential in the vicinity of its maximum value and
therefore describes a very general situation in which the potential
responsible for dark energy has one (or more) maxima.  (In fact Eq.
(\ref{eq:V}) generically describes a dark energy potential near its
maximum value.)  We follow the expansion dynamics of the universe as
the scalar field rolls down this potential and, for a large region in
parameter space, compare the resulting value of the luminosity
distance with observations of high redshift type Ia supernovae. Our
results show that the decaying dark energy model is consistent with
SNe data at reasonably small values of $m$ and $\phi_0$. We see that
the potential declines to half its maximum value on a time scale not
shorter than $\Delta T_{1/2} \simeq 0.6 H_0^{-1}$ ($\Delta T_{1/2}
\simeq 8 \ {\rm Gyrs}$ if $H_0 = 70 \ {\rm km/s/Mpc}$) at the $95.4
\%$ confidence level.

Next, we extend this analysis to the cosine potential (\ref{eq:cos})
which holds the possibility of becoming negative as the field evolves.
The cosine potential has some important new features not shared by
`tracker' quintessence potentials.  Due to its ability to grow more
negative with time, the cosine potential allows for the eventual
cancellation between the positive density of dark matter and the
negative density of dark energy, because of which the universe
collapses at late times.  It is important to note that collapse occurs
generically even in the case of a {\em spatially flat} universe, as
pointed out in \cite{ffkl02}. Comparing our model universe with
supernova observations we determine the time, measured from the
present epoch, to when the universe collapses. At the $95.4 \%$
confidence level, the time to collapse can never be smaller than
$\Delta T_{\rm coll} \simeq 1.3 H_0^{-1}$ if $m \lleq 2$ ($\Delta
T_{\rm coll} \simeq 18 \ {\rm Gyrs}$ for $H_0 = 70 \ {\rm km/s/Mpc}$).

For both dark energy models (\ref{eq:V}) \& (\ref{eq:cos}) our current
accelerating epoch is a transient which ends once the dark energy
density falls below a certain critical value. For the cosine potential
the time until the acceleration of the universe ends ($q(t_0+\Delta
T_{\rm end}) = 0$) is always larger than $\Delta T_{\rm end} \simeq
0.7 H_0^{-1}$, which is about 10 Gyrs if $H_0 = 70 \ {\rm km/s/Mpc}$
(at the $95.4 \%$ CL).  Thus a DDE universe will continue accelerating
for at least 10 Gyrs if $H_0 = 70 \ {\rm km/s/Mpc}$ and $\om \simeq
0.3$.  About 8 Gyrs later, this universe will collapse and head
towards a `big crunch' singularity.

Finally, it is worth pointing out that the decaying dark energy models
discussed in this paper are devoid of horizons, which could be an
attractive feature from the viewpoint of string/M-theory
\cite{sahni02}.

\section{Acknowledgements}

VS acknowledges support from the ILTP program of cooperation between
India and Russia. UA thanks the CSIR for providing support for this
work.  AS was partially supported by the Russian Foundation for Basic
Research, grant 02-02-16817, and by the Research Program ``Astronomy''
of the Russian Academy of Sciences. AS also thanks Profs. J. Narlikar
and V. Sahni for hospitality during his visit to IUCAA when this
project was started.

\end{document}